%% file: main.tex
\documentclass[sigconf]{acmart}
\input{header}

\begin{document}

\title[xGFabric: Coupling Sensor Networks and HPC Facilities with Private 5G Wireless Networks]{xGFabric: Coupling Sensor Networks and HPC Facilities with Private 5G Wireless Networks for Real-Time Digital Agriculture}


\author{
Liubov Kurafeeva,
Alan Subedi,
Ryan Hartung,
Michael Fay,
Avhishek Biswas,
Shantenu Jha,
Ozgur O. Kilic,
Chandra Krintz,
Andre Merzky,
Douglas Thain,
Mehmet C. Vuran,
Rich Wolski
}



\renewcommand{\shortauthors}{Kurafeeva et al.}

\begin{abstract}
\it
Advanced scientific applications require coupling distributed sensor networks with centralized high-performance computing facilities. Citrus Under Protective Screening (CUPS) exemplifies this need in digital agriculture, where citrus research facilities are instrumented with numerous sensors monitoring environmental conditions and detecting protective screening damage. CUPS demands access to computational fluid dynamics codes for modeling environmental conditions and guiding real-time interventions like water application or robotic repairs. These computing domains have contrasting properties: sensor networks provide low-performance, limited-capacity, unreliable data access, while high-performance facilities offer enormous computing power through high-latency batch processing. Private 5G networks present novel capabilities addressing this challenge by providing low latency, high throughput, and reliability necessary for near-real-time coupling of edge sensor networks with HPC simulations. This work presents xGFabric, an end-to-end system coupling sensor networks with HPC facilities through Private 5G networks. The prototype connects remote sensors via 5G network slicing to HPC systems, enabling real-time digital agriculture simulation.
\end{abstract}


\maketitle
\input{intro}

\input{application}

\input{arch-overview}
\input{arch-sensor}
\input{arch-5g}

\input{arch-cspot}
\input{arch-hpc}

\input{arch-e2e}
\input{eval}

\input{related}
\input{conclusion}

\begin{acks}

This work was supported by the US Department of Energy under award DE-SC0025541.
\end{acks}


\bibliographystyle{ACM-Reference-Format}
\bibliography{main}

\appendix
\input{appendix}

\end{document}

%% file: header.tex
\usepackage{graphicx}
\usepackage{pdfpages}
\usepackage{ragged2e}
\usepackage{sc25repro}

\copyrightyear{2025}
\acmYear{2025}
\setcopyright{cc}
\setcctype{by}
\acmConference[SC Workshops '25]{Workshops of the International Conference for High Performance Computing, Networking, Storage and Analysis}{November 16--21, 2025}{St Louis, MO, USA}
\acmBooktitle{Workshops of the International Conference for High Performance Computing, Networking, Storage and Analysis (SC Workshops '25), November 16--21, 2025, St Louis, MO, USA}
\acmDOI{10.1145/3731599.3767589}
\acmISBN{978-8-4007-1871-7/2025/11}

\newcommand{\XGFABRIC} {\textsc{xGFabric}}
\newcommand{\CSPOT} {\textsc{CSPOT}}
\newcommand{\LAMINAR} {\textsc{Laminar}}

\usepackage{xcolor}

%% file: intro.tex
\begin{figure}[t]
    \centering
    \includegraphics[width=\columnwidth]{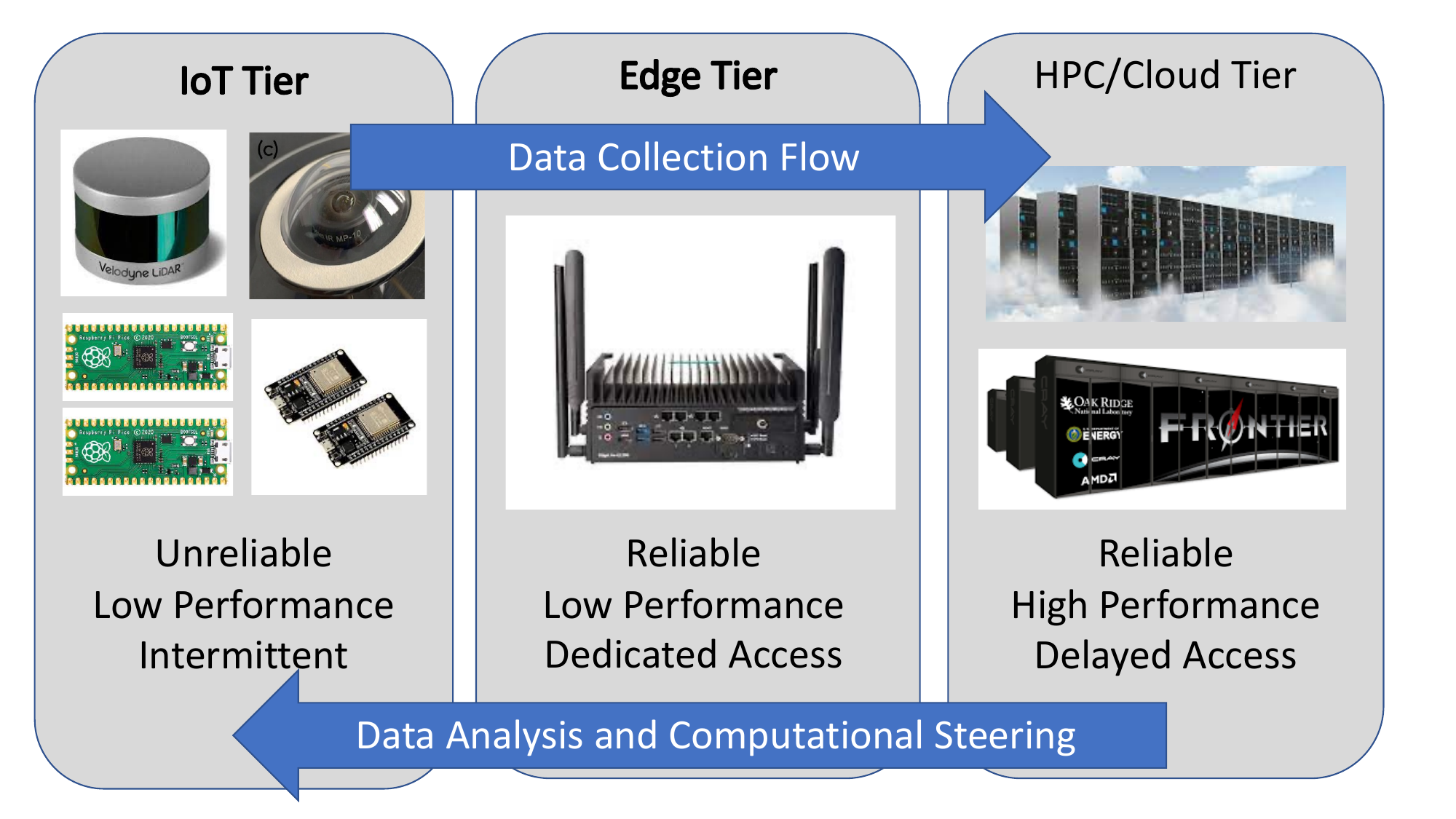}
\caption{End to End Coupling of Sensor Networks and HPC}
\justify
{\it xGFabric provides an end-to-end system for
coupling sensor networks through edge servers to HPC facilities, navigating extraordinary differences in reliability, performance, and
responsiveness across the layers.}
\vspace{-0.4in}
\end{figure}

\section{Introduction}
\label{sec:intro}

The ability to couple large-scale computing facilities with
scientific instruments and sensors (of all scales) so that they can function together as a single system has emerged as a key requirement for new scientific discovery.
Mitigating the effects of climate change on agriculture and
ensuring U.S. energy independence both require modeling and responding to  dynamically changing physical phenomena (e.g. pathogen virulity, propagation and their dependence on external conditions, etc.) that are difficult to predict. While predictions of the relevant phenomena will improve, key to this improvement is the ability to study such phenomena at ever smaller time scales in as close to real time as possible.  

CUPS (Citrus Under Protective Screening) is an
example of a novel digital agricultural application
poised to benefit from the end-to-end coupling
of sensor networks with HPC capability.  CUPS is a new pest-remediation strategy under study by the citrus industry as a sustainable way to protect citrus orchards from huanglongbing (HLB) ``citrus
greening'' disease~\cite{crb-2023}.  Careful monitoring and control of the growing conditions inside a CUPS facility is critical to their commercial success at scale.   CUPS can make use of HPC for large-scale sensor data processing
and for advanced modeling, simulation, and machine learning applications that support sustainable farming practices in this complex setting.

To prototype this concept, we have constructed
\XGFABRIC , a novel distributed system that combines an agricultural sensor network in a facility located in a remote area,
connected by a Private 5G wireless network to the commodity
Internet.  Sensor data from the CUPS is collected, summarized,
and conveyed over the Private 5G network
via the CSPOT (Serverless Platform of Things in C) distributed runtime system,
where it is distributed to HPC facilities at
both campus infrastructure and national computing
facilities.  The arrival of data triggers the
execution of a Computational Fluid Dynamics (CFD) simulation of the airflow and heat transfer inside the CUPS (a 100,000 cubic meter screen house) to predict internal conditions based on sensor measurements at the boundaries.  These results can be returned to the site operator to guide
the application of water, pesticides, or to detect
failures of the protective screening.

We evaluate the capability of the \XGFABRIC\ prototype
in several dimensions.  We measure the performance and capacity of the Private 5G wireless network, the performance
of reliable data delivery via CUPS, and the runtime and speedup
of CFDs on multiple HPC platforms.  We observe that the end-to-end
performance meets the real-time requirements to satisfy the
CUPS facility.

Our contributions are to demonstrate that Private 5G networks offer novel capabilities that can be exploited and extended to provide the low latency, high throughput, and reliability needed to perform the near-real-time coupling of edge sensor networks with simulations running in HPC facilities. Specifically, the coupling of large-scale systems with scientific instruments, sensors, and actuators at all scales requires a new approach to adaptive workflow management and new system software abstractions. We also demonstrate the capabilities of a new software \textit{fabric} that unifies resources at all device scales -- from sensors to large-scale, batch controlled machines -- across different network infrastructures.  This full-stack platform is capable of delivering ``in-the-loop'' high-performance computing capabilities to distributed applications to support decision making in real time.

%% file: application.tex
\begin{figure}[t]
\vspace{-0.5in}
\centering
    \includegraphics[width=\linewidth]{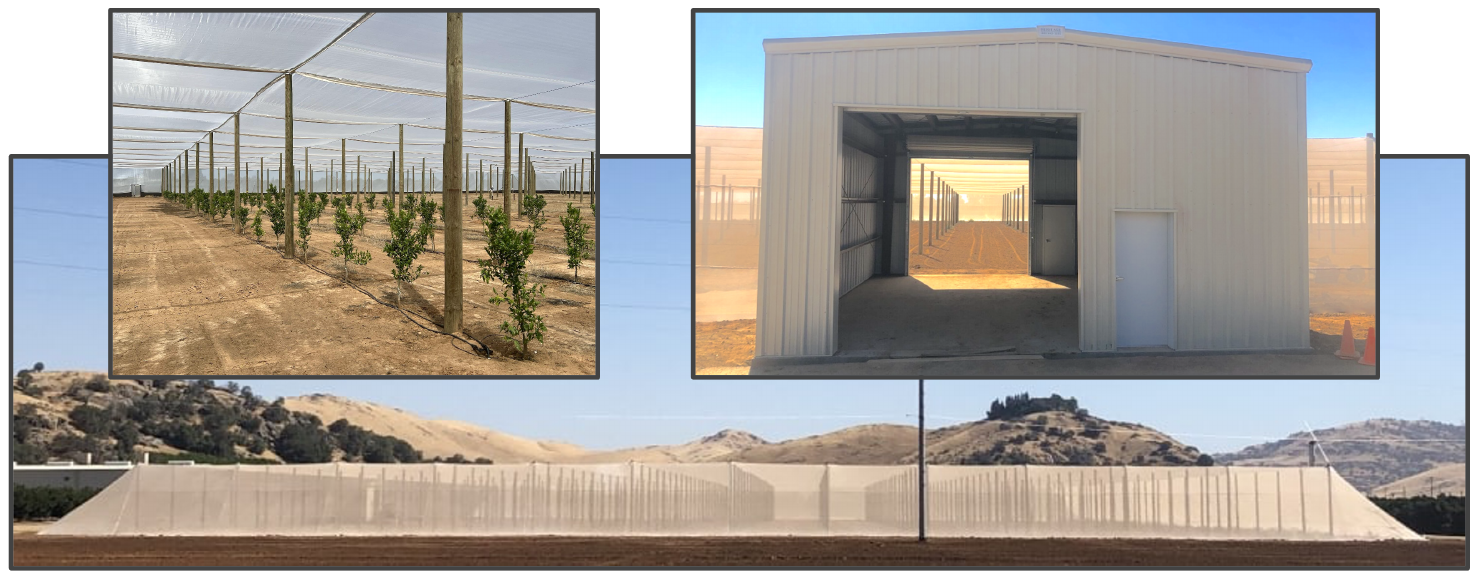}
\vspace{-0.9in}
\caption{Citrus Under Protective Screening}
    \label{fig:cups}
\justify
{\it CUPS is an agricultural research pilot for testing control of {\it huanglongbing} citrus greening disease.  The facility is equipped with a sensor network for detecting local conditions.}
\vspace{-0.2in}
\end{figure}

\section{Application: Citrus Under Protective Screening (CUPS)}
\label{sec:app}

The citrus production industry is currently developing remediation strategies
for the Asian citrus psyllid which carries the huanglongbing (HLB) ``citrus
greening'' disease.  HLB has devastated the commercial citrus industry in
Florida and Texas with an annual cost of more than \$1B US~\cite{crb-2023}.
From a biosafety perspective, HLB is a significant vector.  Pesticides and 
disease-resistant cultivars have, so far, proved ineffective.  
Its effect on citrus production in the south has been rapid and irreversible.

In California, 
where the disease is present but not yet epidemic, growers are experimenting
with siting orchards inside large, protective screen houses.  The Citrus Under
Protective Screening (CUPS) project is an at-scale pilot for screen-house
citrus production located at the Lindcove Research Extension Center in Exeter,
California, shown in Figure~\ref{fig:cups}.  While CUPS is specifically testing HLB control, it represents an
approach that is effective against any insect-born pathogen for which typical
husbandry practices are ineffective.

The goal of CUPS is to understand the growing environment and commercial
agricultural viability of screen-house citrus production.  Citrus trees have
useful production lifetimes that exceed 20 years.  CUPS is effective as long
as the trees that are introduced into the screen house are disease free and
the screen remains in tact.  For commercial viability, the screen houses must
be large (covering several acres each) and they must accommodate tree canopy and
harvesting equipment that require 25 to 30 feet of vertical space.

Detecting and rapidly repairing screen breaches in the commercial scale CUPS is a
critical open problem.  While industrial accidents that cause screen damage
can be detected and rapidly reported by workers, unobserved events (e.g. bird
strike, foraging fauna, damage concomitant with theft, etc.) can cause screen
breeches that must be detected.

Our team has been working to instrument and analyze the growing environment
within the at-scale CUPS structure in Exeter.  As part of that on-going work, 
we have developed
a Computational Fluid Dynamics (CFD) model that
can model to predict airflow within a CUPS screen house in near real-time
based on instantaneous wind, temperature, and humidity measurements taken and
the screen boundaries (both inside and outside).  Analytically, the goal of
the model is to provide growers with decision support for input events such as
pesticide or fertilizer spraying, frost prevention, etc. where the grower must
make a decision regarding timing, location, and quantity of input to apply.

However, we are also exploring whether the model can detect screen breech.
Specifically, once the model is calibrated, a deviation between predicted and
measured airflow can portend a possible screen breech and, perhaps, an area of
the structure where the breech may have occurred.  
Note that we plan to structure the coupling of real-time senor data with CFD
as a ``digital twin'' in which the true atmospheric conditions within the
structure are ``twinned'' by the results of the CFD model for the interior of
the structure.  The model results will inform both modality changes in the
sensing infrastructure and data calibrations (back tested against historical
data) that are necessary to maintain model accuracy.

Our team will also be deploying a Farm-NG~\cite{farm-ng} wheeled robot with 
autonomous-driving capability within the CUPS structure.  As a driver for
\XGFABRIC\ research, our plan is to investigate whether it will be possible to
detect a potential breech (using a large-scale HPC machine to run the CFD
model which is parameterized by real-time \textit{in situ} boundary
conditions), compare the modeled airflow to measurements taken for the same
time period within the structure, and if they do not match, dispatch the robot
to surveil the region of the screen where a breech may have occurred using
an on-board camera.  The \XGFABRIC\ digital-physical fabric will incorporate
robot-based sensing and robot route planning, thereby linking it to, and
augmenting the CFD-based digital twin for the screen structure.  

This ambitious application illustrates how a digital-physical fabric can
enable new biosecurity capabilities.  However, to bring it to fruition
requires the ability to amalgamate computational resources at all scales, and
to ``close the loop'' between sensing, computing and storage, and actuation.

%% file: arch-overview.tex
\begin{figure*}[t]
    \includegraphics[width=0.9\textwidth,trim={0 2.25in 0 0},clip=true]{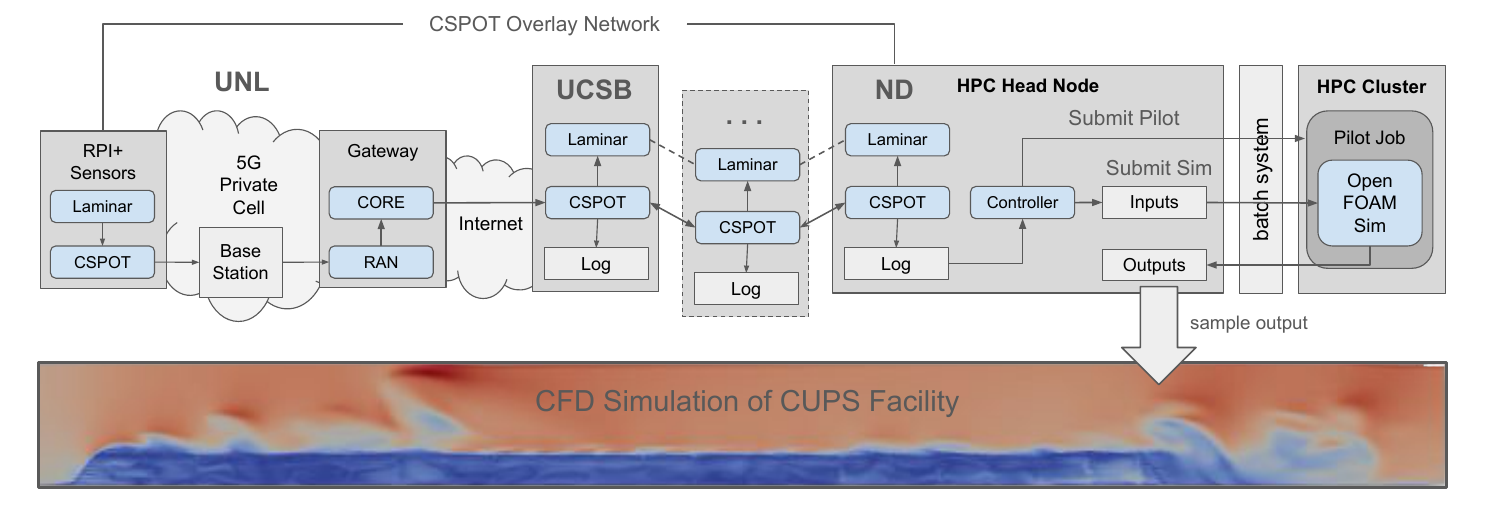}
    \vspace{-0.2in}
    \caption{\XGFABRIC\ Architecture and Sample Output}
        \label{fig:architecture}
\justify
{\it xGFabric connects a remote sensor network (left) with an HPC facility (right).
In this prototype, a sensor network at UNL (U. Nebraska-Lincoln) consists of Raspberry Pis running the CUPS distributed runtime system.  These are connected by a private 5G network to the commodity Internet, and communicate with a network of CUPS nodes at UCSB (U. California - Santa Barbara), ND (U. Notre Dame),
and other facilities.  At ND, the Controller process dispatches a pilot
job, constructs the input data for OpenFOAM, and runs the CFD code when
sufficient sensor data arrives. }
\vspace{-0.1in}
\end{figure*}

\section{XGFabric Architecture}
\label{sec:arch}

The \XGFABRIC\ architecture is shown in Fig.~\ref{fig:architecture}. To the best of our knowledge, \XGFABRIC\ is the first end-to-end distributed system to seamlessly \textbf{integrate field wireless sensor networks with high-performance computing (HPC) workflows in real time}. This integration results from a full-stack, multi-scale software approach unifies devices using a private 5G wireless network architecture, with edge, cloud, and HPC systems using the the \CSPOT~(Serverless Platform of Things in C)~\cite{wolski2019cspot} distributed runtime and the the \LAMINAR\ ~\cite{laminar} dataflow  system.

\subsection{Overview}
\label{sec:arch:overview}

Note that the \XGFABRIC\ software stack (described below) is designed to run natively (on microcontrollers), as a runtime using Linux, or as a containerized guest.  Thus it is complementary to large-scale deployment infrastructure such as Sage~\cite{sage,sage-web} which includes Waggle~\cite{waggle} as a software stack, or the Array of Things~\cite{array-of-things}.  It differs from these approaches in that it is full-stack (including a network-transparent dataflow programming environment), it includes support for managing HPC workloads, and it targets 5G/6G wireless infrastructure at the edge.  Interfacing \XGFABRIC\ to Sage via Waggle is the subject of our future research efforts.

The \XGFABRIC\ architecture uses the \CSPOT\ log-based, distributed event system to implement reliable, delay-tolerant networking, end-to-end, from devices in a private 5G network to a batch-controlled HPC machine and vice versa. \XGFABRIC\ leverages this delay tolerance in three ways: first, devices operating in remote locations using 5G connectivity can be subject to frequent network interruption.  Because \textit{all} program state is logged, programs can simply pause until connectivity is restored.  Secondly, \CSPOT\ logs are implemented in persistent storage, so power-loss (which is frequent in remote IoT settings) and other device failures that do not destroy the log storage are treated in the same way as network interruption.  Since all program state updates are implemented as log appends, a ``failure to append'' to some program log, which results either from network interruption or node failure, is simply retried until it succeeds or the application terminates the computation. Third, \XGFABRIC\ uses this delay-tolerance to 
mask batch-queing delays on HPC systems that are batch-controlled.  Data is
``parked'' in logs on storage accessible by the compute nodes of a cluster and
fetched once the nodes become active.

In the following, we provide details on each of the \XGFABRIC\ components: the sensor network, the private 5G wireless network, the \CSPOT\ distributed runtime system, the \LAMINAR\ data flow language, the HPC interface, and
finally the end-to-end operation.

%% file: arch-sensor.tex
\subsection{Remote Sensor Network}
\label{sec:arch:sensor}

The sensor network layer in xGFabric consists of edge devices used to collect, pre-process, and transmit physical-world data in real time. These devices connect exclusively through a private 5G cellular network, which provides the uplink channel for transmitting data to centralized compute or storage infrastructure.
The edge devices primarily include Raspberry Pi 4 units equipped with 5G USB modems. Each unit runs a software agent called CSPOT, which continuously forwards sensor data using standard IP networking protocols to external endpoints. 
The system supports multiple concurrently connected user equipments (UE), each transmitting independently. The use of a private 5G network enables precise control over radio resources and performance at the edge.
This sensor network forms the entry point into xGFabric’s virtualized data collection pipeline, supporting modular, replicable deployments across multiple locations and ensuring high availability and robust wireless connectivity.

%% file: arch-5g.tex
\vspace{-0.1in}
\subsection{Private 5G Wireless Network}
\label{sec:arch:5g}

5G networks provide the connectivity needed to access
sensor networks in remote locations.
Network slicing~\cite{Batur23Informs}, a key capability of 5G, enables the creation of multiple virtual networks slices within the same physical infrastructure, each tailored to different application demands. This allows the network to simultaneously support diverse use cases such as low-latency control systems, high-throughput video, or lightweight IoT traffic.

For the \XGFABRIC\ prototype we deploy two private 5G wireless networks that support both development and production environments, built using the open-source srsRAN\cite{srsran} stack and Open5GS\cite{open5gs} core for standalone 5G functionality and a custom-made CI/CD workflow.
The underlying hardware architecture centers around a single compute device that hosts both the development and production private 5G network functions. The device is equipped with an Intel Core i7 processor, 32 GB of DDR4 memory, and 1 TB of solid-state storage, running Ubuntu 24.04 LTS. Network connectivity is handled via a high-performance Intel 82599ES 10-Gigabit Ethernet network interface card, and wireless capabilities are supplemented with onboard Wi-Fi. Two software-defined radios (SDRs)—a USRP B210~\cite{ettus} and a USRP B200~\cite{ettus} from Ettus Research—serve as the RF frontends for the two private 5G networks. These SDRs are connected to an OctoClock~\cite{ettus} timeserver to enable precise time synchronization across the system.

Both the development and production private 5G networks are deployed using Docker~\cite{docker} containers. Each network instance includes a gNodeB (gNB) component that interfaces with its corresponding SDR and handles radio access network (RAN) operations, including scheduling, modulation, and UE signaling. Moreover, each instance runs a containerized 5G core network stack using Open5GS~\cite{open5gs}, which provides a full suite of standalone (SA) 5G core functionalities (i.e., subscriber authentication, session and mobility management, policy enforcement, and data routing).

The development and production private 5G instances utilize different sets of UEs for testing and validation. In the development instance, we connect a Google Pixel 6a commercial off-the-shelf (COTS) smartphone and two Raspberry Pi 5 devices, each configured with an RM530N-GL 5G USB modem~\cite{5gDongle}. In the production instance, we connect two Raspberry Pi 4 units, each paired with its own RM530N-GL dongle. All UEs rely on programmable sysmoISIM-SJA5~\cite{5gsim} SIM cards for registration and authentication with the 5G core network. These SIM cards are provisioned using the open-source pysim~\cite{pysim} toolkit, allowing for flexible and consistent identity management across both environments.

By supporting two parallel private 5G networks within the same physical infrastructure, we ensure flexibility in experimentation and deployment. The development instance allows for safe testing of new features such as network slicing, while the production instance maintains a consistent baseline for evaluating performance, reliability, and multi-UE scenarios.
This architecture provides a foundation for continuous innovation in private 5G and future 6G systems, while preserving stability and reliability in production scenarios. The evaluation results below are acquired from the production environment.


%% file: arch-cspot.tex
\vspace{-0.15in}
\subsection{\CSPOT\ Distributed Runtime System}
\label{sec:arch:cspot}

\CSPOT\ is a distributed runtime system that provides
reliable multi-node communication built on log based storage.
It  is designed function at all device
scales, from microcontrollers to edge-based computers to large-scale HPC and cloud systems.
It uses logs (which are simple to implement efficiently at all scales) as
persistent program variables.  As a result, a \CSPOT\ program can be
interrupted at any moment and the current program state will be available in
persistent storage so that the program can be immediately resumed after the
interruption.

Another feature of the log-based event system that underpins \XGFABRIC\ is that
it is highly concurrent.  In particular, only the assignment of a unique sequence
number with a specific log entry appended to a log must be implemented
atomically.  Logs are otherwise accessible concurrently.  Further, to obviate
the need for lock-recovery for locks that span network connections, \CSPOT\
does not include a lock function as part of its API.  Internally, the \CSPOT\
implementation uses locks to implement atomic sequence number assignment, but
it does so in a way that prevents locks from being held while a thread is
waiting for network communication. 

As a result, a \CSPOT\ append operation fails in only one of two ways.  Either the append fails, and the API call returns an error, or the append succeeds but the sequence number associated with the append (to be returned from the API call) is lost, generating an error.  Retrying the append until a sequence number is successfully returned ensures data integrity, but deduplication of the \CSPOT\ logs is necessary to implement ``exactly once'' delivery semantics. This feature greatly enhances crash
resilience and network partition tolerance, but it makes \CSPOT\ non-intuitive
for developers familiar with more conventional concurrency management
mechanisms.

In particular, there is no way in a \CSPOT\ program to fire an event only after
multiple appends (to the same or different logs) have occurred.  Event
handlers are the only computational mechanism and a handler can only be
triggered by a single log append (to avoid locking by handlers waiting for
future events).  As a result, a \CSPOT\ program can always make progress (no
handler blocks waiting for another handler) but handler code must parse and
scan the logs to implement multi-event synchronization.  

\vspace{-0.15in}
\subsection{\LAMINAR\ Dataflow System}

To improve programmability over using logs and events as native programming abstractions, \XGFABRIC\
includes a distributed dataflow~\cite{dataflow} programming environment,
called \LAMINAR, that hides \CSPOT\ synchronization complexity within a
relatively conventional dataflow framework.  \LAMINAR\ implements a
strongly-typed applicative language with strict~\cite{df-strict} semantics
using \CSPOT\ as its runtime system.  Note that \CSPOT's implementation of
logs makes each log a ``single-assignment'' variable from the programming
language perspective.  Thus it is possible to implement functional programming
semantics (such as strict, applicative dataflow) using \CSPOT.
As a result, \LAMINAR\ shares \CSPOT's failure
resiliency and crash-consistency properties~\cite{laminar-fmec} while
implementing (on behalf of the programmer) many of the optimizations needed to
avoid log scans during synchronization. 

While \LAMINAR\ is strongly-typed, it allows the developer to specify
application-specific types.  Thus any computation that produces the same
outputs from a given set of inputs (e.g. any ``stateless'' computation) can be
embedded within a \LAMINAR\ computational node and ``fired'' as part of a
\LAMINAR\ dataflow program.  For example, it is possible to treat a large-scale
Computational Fluid Dynamics (CFD) application as a single node within an encompassing
\LAMINAR\ program that handles the CFD inputs and outputs.

Note that \CSPOT\ and \LAMINAR\ are network transparent and support multiple
network protocols within the same application deployment.  As such, they manage
the network fabric on behalf of the application.  Thus, an \XGFABRIC\
application need not interact with the 5G network configuration interface
directly.  Instead, they operate a network slicing interface to
configure the private 5G network according to the connectivity needs of a specific
deployment.    

%% file: arch-hpc.tex
\subsection{HPC Pilot Interface}
\label{sec:arch:hpc}

\XGFABRIC\  uses the Pilot~\cite{pilot} mechanism from Radical-Cybertools to dynamically configure the HPC environment for large-scale parallel computations. The aim of the pilot and its scheduling/placement algorithm is to bridge real-time data flows with HPC simulations. Interactive pilots ensure rapid responsiveness, ideal for real-time tasks, whereas batch pilots optimize throughput and resource utilization for compute-intensive tasks at the cost of latency from scheduling.
The Pilot Controller currently initiates an initial pilot using a single node and employs the following decision logic to dynamically allocate resources:

\begin{enumerate}
\item Assess incoming data size \(D\) and choose nodes \(N_{req}\):
\begin{equation}
N_{req} = \max\left(1, \frac{D}{threshold}\right)
\end{equation}

\item Evaluate currently available nodes \(N_{avail}\):
\begin{equation}
N_{avail} = \sum_{p \in \text{active pilots}} nodes(p)
\end{equation}

\item Decide whether to submit a new pilot:
\begin{equation}
\text{Submit Pilot} = \begin{cases}
\text{No}, & N_{avail} \geq N_{req} \\
\text{Yes}, & N_{avail} < N_{req}
\end{cases}
\end{equation}

\item Determine pilot submission parameters:
\begin{equation}
nodes = \min(\text{system nodes}, N_{req}),
\end{equation}
\begin{equation} 
\quad runtime = \min(\text{max system runtime}, \text{estimated task runtime})
\end{equation}
\end{enumerate}


As future work, we plan to explore proactive (\textit{starting pilots early}) and reactive (\textit{starting pilots on-time}) strategies to further enhance system responsiveness and efficiency. Proactive pilots reduce latency but may incur idle resource overhead, while reactive pilots minimize idle resources but can introduce startup delays.


%% file: arch-e2e.tex
\subsection{End-to-End Operation}
\label{sec:arch:e2e}

In Fig.~\ref{fig:architecture}, \XGFABRIC\ is depicted in the context of a working,
end-to-end application that dynamically triggers a CFD computation in response
to changing localized atmospheric conditions in an agricultural setting.

As shown in Fig.~\ref{fig:architecture}, atmospheric measurements from sensors are gathered through a private 5G network at UNL, and relayed to a data repository located at UCSB using native \CSPOT. In
addition, a \LAMINAR\ program distributed between UNL and UCSB monitors the
telemetry stream to detect when conditions change.  The measurement errors
from the atmospheric sensors (commodity commercial agricultural weather stations) are
high enough so that consecutive readings may not be statistically determinable
to be
``different'' and, thus, an updated CFD calculation would potentially waste
HPC resources computing a new result that is statistically indistinguishable
from the previous result.  When the \LAMINAR\ change-detection program
determines that conditions have meaningfully changed, it triggers the Pilot to launch a new
CFD computation on the HPC machine located at ND.  The Pilot gathers the most recent atmospheric 
telemetry from the \CSPOT\ logs at at UCSB and launches a preprocessing
pipeline to generate input files and meshing coordinates for the CFD
computation (which is implemented using OpenFOAM~\cite{openfoam}).  Once these
files have been prepared, the Pilot launches the computation in the ND batch
queue and waits for the results to be generated as a set of output rasterized files.

%% file: eval.tex
\section{Evaluation}
\label{sec:eval}

We evaluate the individual elements of \XGFABRIC\ to demonstrate that performance goals are achieved for
the private 5G network, the communication latency of the \CSPOT\ sensor network, and the response time of the
OpenFOAM simulation code.

\vspace{-0.05in}
\subsection{Private 5G Wireless Communications}
\label{sec:eval:5g}

To evaluate the performance of the private 5G network, we conduct three experiments. First, we measure single-user uplink throughput across varying bandwidths, duplexing modes, and device types. Second, we extend the setup to a two-user scenario to assess simultaneous uplink performance. Third, we evaluate the effect of network slicing on throughput by configuring two user equipments on distinct slices with complementary PRB allocations. The single-user and two-user uplink tests are also compared to results from our previously deployed private 4G wireless network.

\begin{figure}[t]
     \centering
    \includegraphics[width=0.9\linewidth]{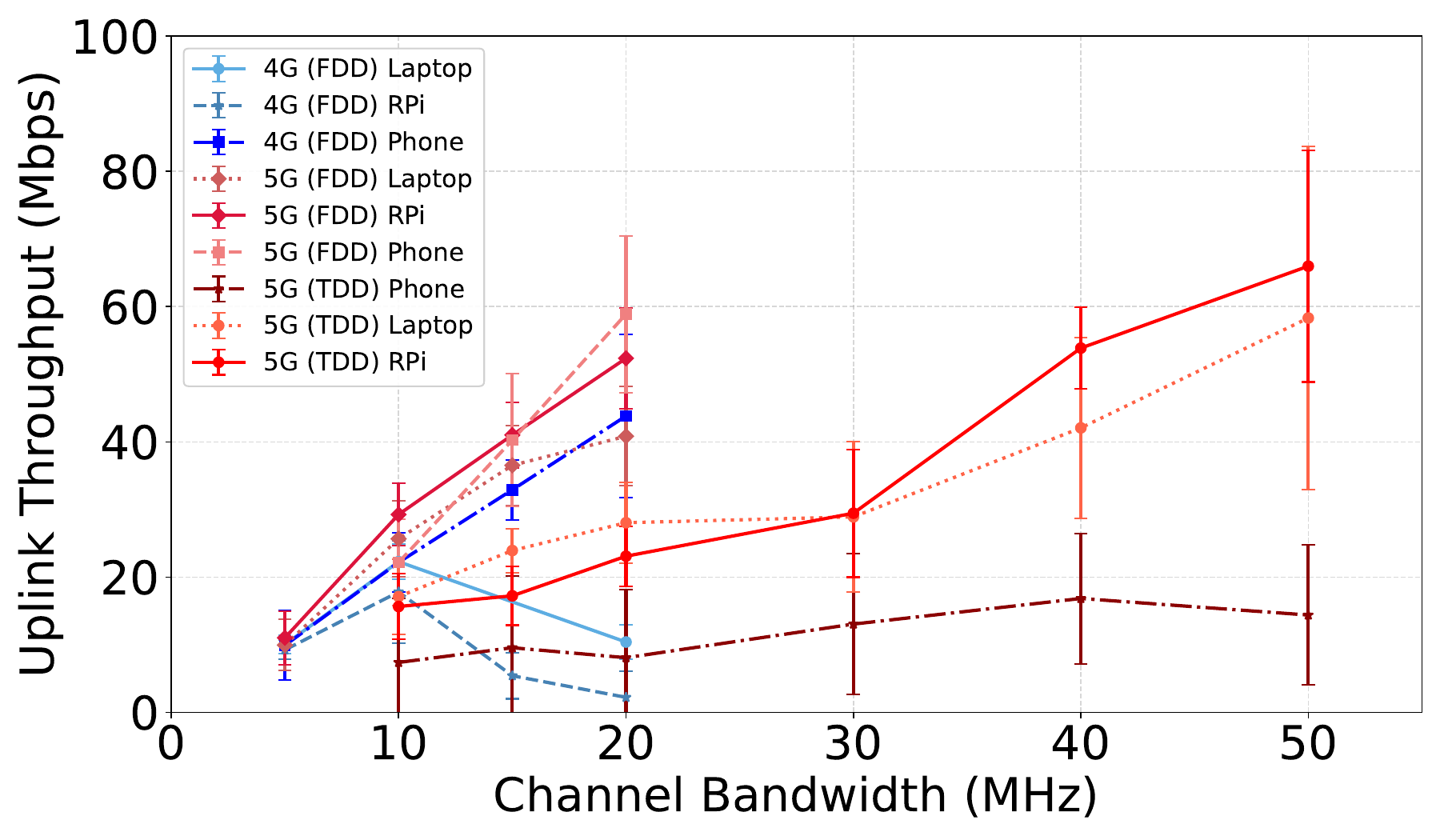}
    \vspace{-0.1in}
    \caption{Single-user Uplink Throughput Across Devices}
    \label{fig:singleUE}
    \vspace{-0.2in}
\end{figure}

In Fig.~\ref{fig:singleUE}, it is shown how uplink throughput performance varies across different bandwidths, duplexing modes, and user devices in a single-user scenario. First, a laptop connected to a 4G Frequency Division Duplex (FDD) network using the SIM7600G-H 4G modem~\cite{4gDongle} is tested. Starting at 5 MHz, the bandwidth is progressively increased to 10, 15, and 20 MHz, while collecting 100 iperf3 uplink throughput samples at each step. The same experiment is then repeated using a Raspberry Pi (RPi) with the same 4G modem, and again with a commercial smartphone.
For the 5G experiments, a laptop equipped with the RM530N-GL 5G modem is connects to the 5G FDD network, and an uplink throughput is measured at bandwidths 5MHz, 10MHz, 15MHz, and 20MHz, with 100 iperf3 samples are collected at each step. The same procedure is repeated using a Raspberry Pi keeping the same 5G modem, followed by a commercial smartphone.
Next, experiments are conducted on a 5G Time Division Duplex (TDD) network. A laptop with the RM530N-GL modem is connected and tested at bandwidths 10MHz, 15MHz, 20MHz, 30MHz, 40MHz, and 50MHz, collecting 100 samples at each setting. This is repeated using the RPi and the smartphone. These experiments allow a comprehensive comparison of how bandwidth, duplexing mode, and device type influence uplink throughput in private cellular networks.

The results in Fig.~\ref{fig:singleUE} demonstrate that uplink throughput scales with bandwidth but is significantly influenced by device type and duplexing mode. In 4G FDD network, the smartphone achieves the highest throughput at 20 MHz (43.83 Mbps), outperforming both the laptop (10.41 Mbps) and the RPi (2.23 Mbps). The limited performance of the laptop and RPi beyond 10 MHz is likely due to constraints imposed by the external 4G modem used in these setups. In 5G FDD network, all devices show marked improvement, with the smartphone reaching 58.89 Mbps, the RPi 52.36 Mbps, and the laptop 40.83 Mbps. In 5G TDD network, the RPi achieves the highest overall throughput (65.97 Mbps at 50 MHz), outperforming both the laptop (58.31 Mbps) and the smartphone (14.40 Mbps). Throughput variability increases with bandwidth, particularly in TDD mode. Overall, while smartphones lead in 4G, laptops and RPis offer competitive—and in TDD, superior—performance in 5G networks when paired with capable modems.

\begin{figure}[t]
    \centering
    \includegraphics[width=0.95\linewidth]{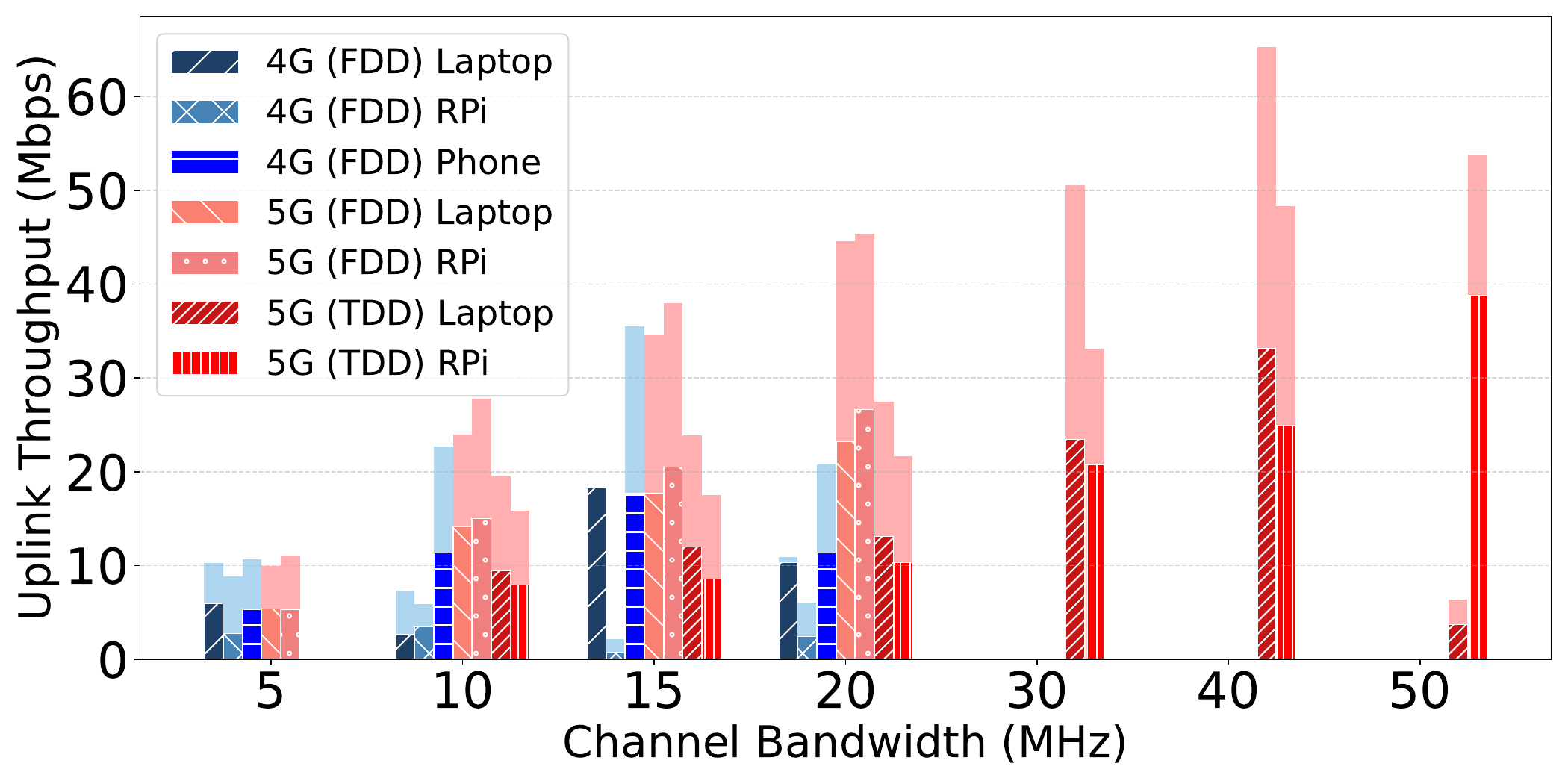}
    \vspace{-0.1in}
    \caption{Two-user Uplink Throughput Across Devices}
    \label{fig:multiUE}
    \vspace{-0.2in}
\end{figure}

In Fig.~\ref{fig:multiUE}, we show how the uplink throughput performance varies across different channel bandwidths, duplexing modes, and user devices in a two-user scenario. First, two laptops equipped with SIM7600G-H 4G modems are connected to a 4G FDD network operating at 5MHz channel bandwidth. Both devices simultaneously perform iperf3 uplink tests, and 100 throughput samples are collected. The same experiment is repeated at 10, 15, and 20 MHz bandwidths. This process is then repeated using two Raspberry Pis with the same 4G modems, followed by two commercial smartphones.
For the private 5G network, experiments are first conducted in FDD mode. Two laptops equipped with RM530N-GL 5G modems are connected to the 5G FDD network, and simultaneous iperf3 uplink tests are performed at 5, 10, 15, and 20 MHz channel bandwidths, with 100 samples collected at each setting. The same procedure is then repeated using two Raspberry Pis with the same 5G modem, followed by two commercial smartphones.
Next, the same set of experiments is conducted on a 5G TDD network. Two laptops equipped with RM530N-GL modems are first connected to the TDD network, and simultaneous iperf3 uplink tests are performed at 10, 15, 20, 30, 40, and 50 MHz bandwidths. The experiment is then repeated using two Raspberry Pis with the same 5G modem, and finally, two commercial smartphones.

In the two-user scenario, similar to the single-user case, throughput distribution and scaling vary notably across devices and network types. On the 4G FDD network, smartphones scale well up to 15 MHz—reaching 35.5 Mbps—before dropping at 20 MHz, likely due to SDR sampling constraints. Laptops peak at 36.1 Mbps at 15 MHz but show uneven user allocation, while Raspberry Pis degrade with bandwidth due to 4G modem limitations.
In the 5G FDD network, laptops scale from 9.9 Mbps to 45.7 Mbps with balanced performance. Raspberry Pis achieve similar results, peaking at 45.4 Mbps at 20 MHz with fair sharing. The 5G TDD network offers strong scalability at wider bandwidths: laptops reach 65.2 Mbps at 40 MHz before dropping at 50 MHz due to SDR limitations, while Raspberry Pis peak at 53.8 Mbps. Both FDD and TDD modes deliver high and evenly distributed uplink throughput, with TDD supporting broader bandwidth scaling and FDD demonstrating strong, reliable performance within its operational range.

\begin{figure}[t]
    \centering
    \includegraphics[width=0.9\linewidth]{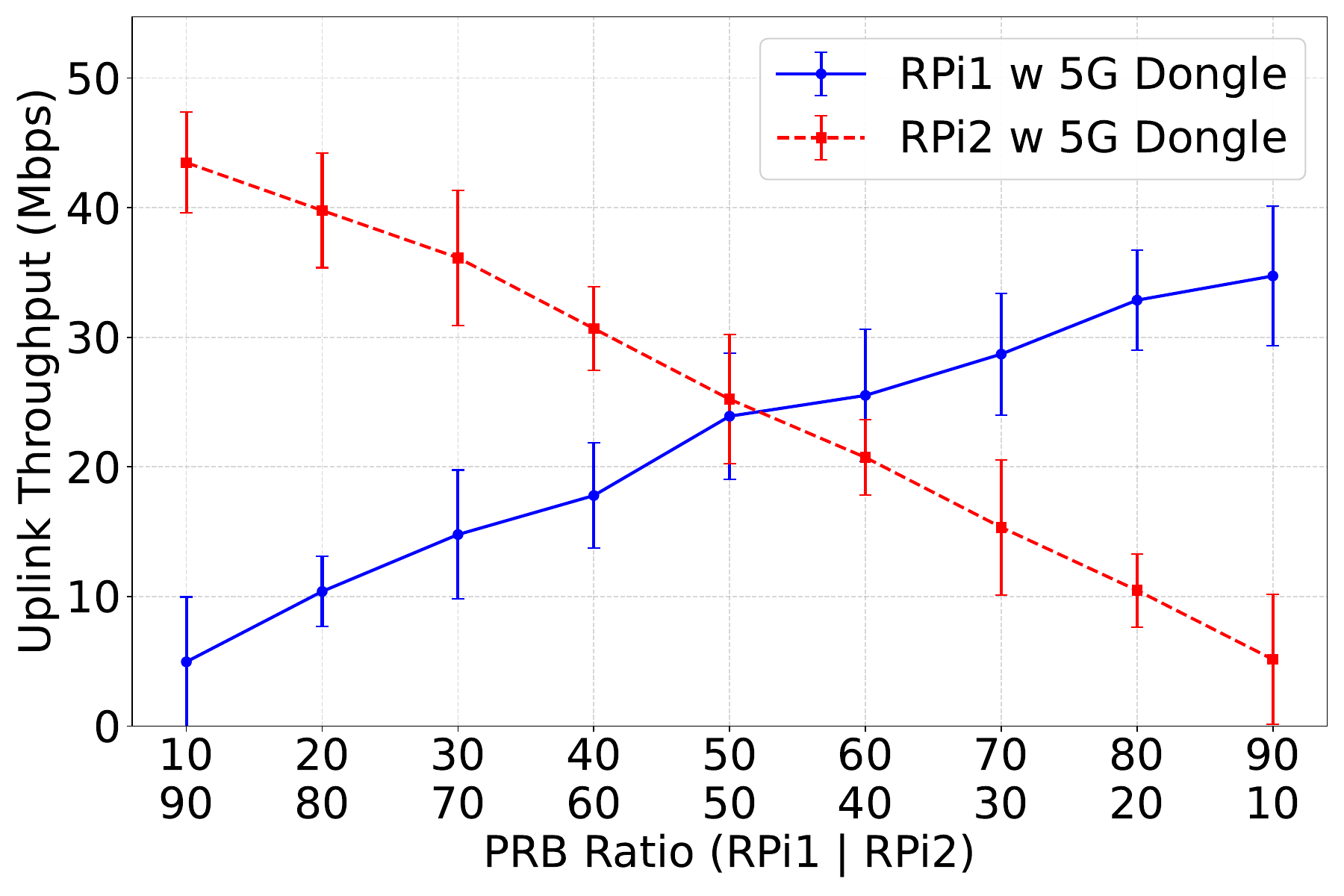}
        \vspace{-0.1in}
    \caption{Two-user Uplink Throughput on a 40 MHz Private 5G TDD Network With Varying PRB Slice Ratios}
    \label{fig:slicing}
    \vspace{-0.2in}
\end{figure}
In Fig.~\ref{fig:slicing}, we show the results of slicing experiment conducted on the private 5G TDD network with a total bandwidth of 40MHz. Two raspberry pi's equipped with RM530N-GL 5G modems are simultaneously connected to the network, each assigned to a different network slice. The system is configured with nine slice profiles, where each slice corresponds to a fixed allocation of physical resource blocks (PRBs), the fundamental units used to allocate radio frequency spectrum in 5G. These slices range from 10\% (slice 1) to 90\% (slice 9) of the total available PRBs.
In the first configuration, RPi1 is assigned to slice 1 (10\% PRBs), while RPi2 is assigned to slice 9 (90\% PRBs). An iperf3 uplink throughput test is performed simultaneously on both devices, and 100 samples are collected per device. In the next configuration, Rpi1 is assigned to slice 2 (20\%), and RPi2 to slice 8 (80\%). This pattern continues with RPi1 progressing from slice 1 to slice 9, and RPi2 in reverse from slice 9 to slice 1, maintaining complementary PRB ratios that always sum to 100\%.

The results also show a clear correlation between PRB allocation and uplink throughput. As each Raspberry Pi is assigned a larger share of the network slice, throughput increases consistently. RPi1 achieves 4.95Mbps at 10\% PRB allocation and scales up to 34.73Mbps at 90\%, while RPi2, assigned the complementary share in each configuration, increases from 5.14Mbps to 43.47Mbps. Midpoint allocations, such as 50\%, yield comparable results—RPi1 and RPi2 achieve 23.91Mbps and 25.22Mbps, respectively. Standard deviations remain within a narrow 3–5Mbps range, indicating stable performance across all slice levels. Throughput generally scales in proportion to the assigned PRBs, demonstrating that the slicing configuration effectively partitions radio resources.
Overall, the experiment confirms that network slicing enables controlled and predictable resource allocation in the private 5G network.

\subsection{\CSPOT\ Sensor Network}
\label{sec:eval:cspot}

From a performance perspective, end-to-end, the application consists of two data paths.  The first transmits telemetry data to be used to determine the initial conditions of the CFD simulation from UNL to UCSB every $5$ minutes.  This $5$-minute interval is the reporting interval of the weather stations deployed in and around the CUPS.
On the second data path, a Laminar program reads the most recent $6$ telemetry values (covering the most recent $30$ minutes) and compares them to the previous $30$-minute period using three different tests of statistical difference.  If conditions have changed in a way that is statistically measurable under the assumptions of the tests, it generates an alert indicating that a new CFD simulation is needed.  The alert status is stored in a \CSPOT\ log at UCSB and fetched to ND on a $30$-minute duty cycle.  The Laminar program components can be deployed either within the private 5G network or at UCSB in any combination.  We execute the statistical tests and a voting algorithm to arbitrate between them at UCSB in this study.

Because both the telemetry data path and the Laminar data path use \CSPOT\ as a message transport, we report the \CSPOT\ message performance for the prototype.
We measure the time to deliver 1 $1KB$ message payload,
30 times back-to-back. (The first of $30$ measurements is discarded because of the initial connection start-up penalty.)  Further, each message is acknowledged with a sequence number after the data has been appended to a log in persistent storage.  Table~\ref{tab:cspot-message} shows the average latency for delivering a $1$KB message payload to persistent storage at the end of a log.

\begin{table}[t]
  \centering
  \caption{\CSPOT\ Message Latency for $1$KB payload.\label{tab:cspot-message}}
  \vspace{-0.1in}
  \begin{tabular}{lrr}
    \toprule
    Path & Latency Avg. (ms) & Latency SD (ms) \\
    \midrule
    UNL->UCSB (5G+Int.) & 101 & 17 \\
    UNL->UCSB (Internet) & 17 & 0.8 \\
    UCSB->ND (Internet)  & 92 & 1 \\
    \bottomrule
  \end{tabular}
  \vspace{-0.25in}
 \end{table}

Three configurations are measured.
`UCSB->UNL (5G+Int.)` measures the latency from a client at
UNL carried over the 5G network and the public internet
to the \XGFABRIC\ node at UCSB.  `UNL->UCSB (Internet)`
is the same measurement, but moving the client to
a wired Ethernet connection to the Internet, skipping
the 5G wireless network.  `UCSB->ND` measures between
CSPOT nodes at UCSB and ND over the public Internet.

Note that the current \CSPOT\ internal messaging protocol (which uses ZeroMQ~\cite{zmq} as a transport) is optimized for reliability and not message latency.  For example, to append data to a remote \CSPOT\ log requires the client to request the size of a log element (which is fixed for each log and stored with the log as part of its header) from the site where the log is hosted before the data is actually sent from the client to the log.  This size is used to determine the size of the message sent from the client carrying the element to be appended.  Earlier versions of \CSPOT\ focused on message latency used caching of the element size on the client side to avoid fetching the element size each time.  This optimization effectively halves the message latency, but causes the append to fail if the log element size is changed on the server side without a client cache update.

While these and other optimizations are possible, for the prototype \XGFABRIC\ application where new telemetry data is available every $300$ seconds and change-detection is performed by the Laminar program every $30$ minutes, further reducing the message latency by as much as an order of magnitude will not appreciably affect application performance.  For example, the effect of moving the telemetry sources from the private 5G network (latency $101$ms) to the Internet directly ($17$ms) -- an order of magnitude improvement in message latency -- would be imperceptible end-to-end.

This result shows that the current production CUPS deployment, that uses a combination of $900$MHz and long-distance Wi-Fi connectivity in and around the CUPS, could be replaced by a private 5G network without ill effect.  Doing so, in the future, will obviate the current solar and battery power distribution infrastructure, thereby drastically reducing the maintenance cost.

\vspace{-0.1in}
\subsection{HPC Simulation Portability}
\label{sec:eval:hpc}

Future deployments of \XGFABRIC\ will make use of
varying HPC sites in order to exploit the changing
availability and performance of different facilities.
To that end, the simulation was successfully deployed and evaluated across three distinct HPC environments: Notre Dame's Center for Research Computing (CRC), Purdue's ANVIL, and the University of Texas' Advanced Computing Center's (TACC) Stampede3. This multi-site deployment strategy enabled assessment of portability challenges and performance characteristics across heterogeneous facilities.

Some practical differences between sites are easily
observed, such as operating system and batch scheduler.  Anticipating these and future
differences requires developing scripts that perform
various checks, resource allocation specifications, and user prompts within the scripts for each computing environment, along with the use of Miniconda to capture
and deploy Python components.   This strategy ensures reproducible builds by maintaining explicit version specifications for all packages and libraries. 

The primary portability challenge emerged from variations in pre-installed software modules across the computing sites. Each HPC system provided different versions of OpenFOAM and ParaView with distinct dependency requirements and compilation configurations. The ParaView installations varied in their graphics library dependencies. This heterogeneity created several issues when creating display environments for rendering the VTK output files generated by the OpenFOAM simulations.

To resolve the visualization rendering challenges, a front-end SSH-based solution was implemented that requires users to establish display-forwarded connections to head nodes for offscreen rendering. While batch job submission with embedded environment variables represents an alternative approach, the complexity of dynamically detecting graphics library configurations and available virtual display systems across diverse HPC environments proved prohibitive. Specifically, Notre Dame and ANVIL systems utilized OpenGL-compiled ParaView with X.Org display servers supporting virtual framebuffer allocation, while Stampede3 employed Mesa-compiled ParaView. ANVIL's configuration presented additional constraints, lacking support for both virtual framebuffer and Mesa environment pass-through capabilities.

\begin{figure}[t]
    \centering
    \includegraphics[width=0.8\linewidth] {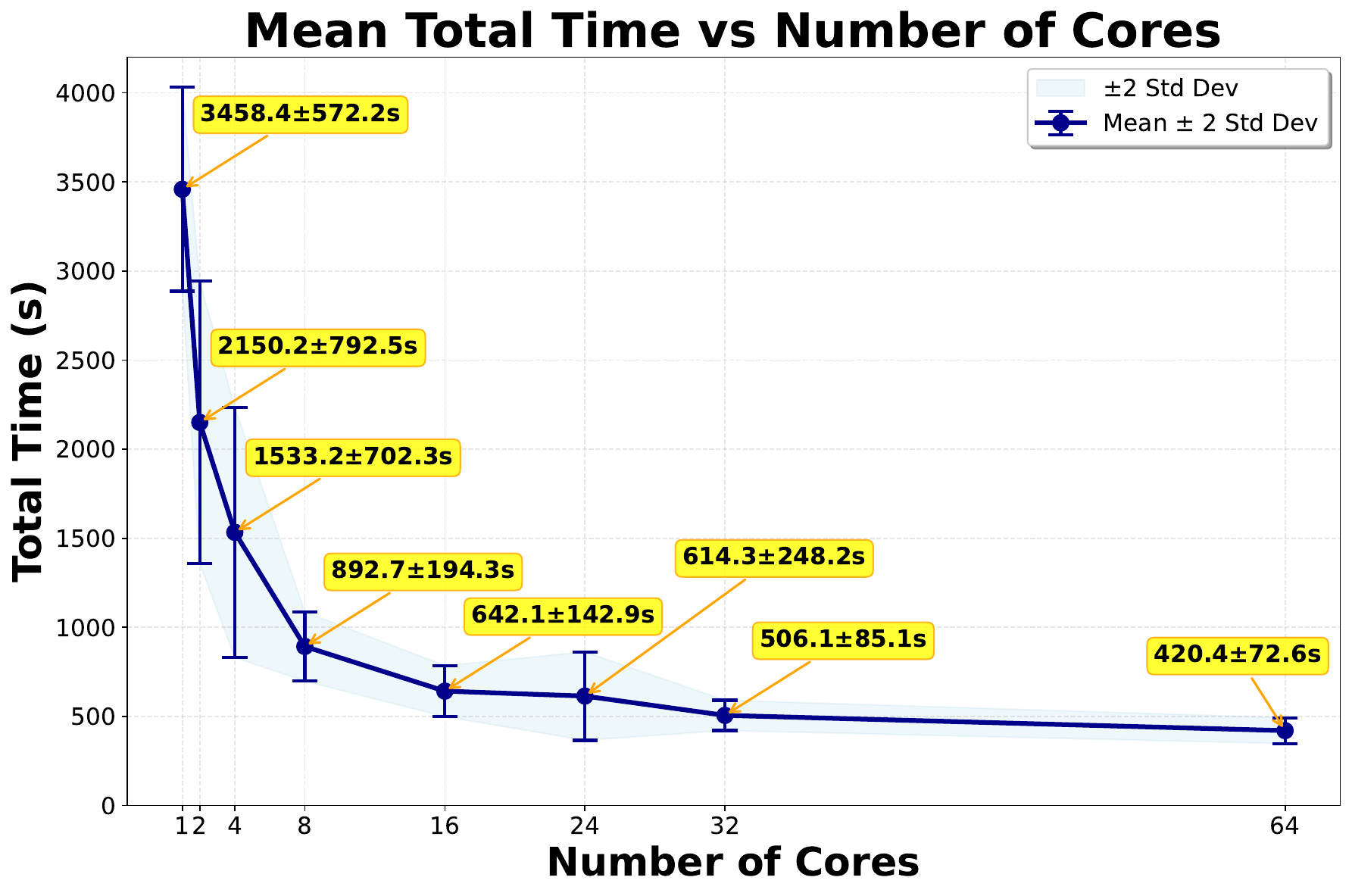}
        \vspace{-0.1in}
    \caption{OpenFOAM Performance}
\justify
{\it Single-node speedup curve for OpenFOAM simulation on 64-core single
node, 10 runs per core count, mean and 2 standard deviations shown.}
    \label{fig:speedup-single}
    \vspace{-0.20in}
\end{figure}

Computational performance remained relatively consistent across all three deployment sites.
Fig.~\ref{fig:speedup-single} shows the performance of full CFD computation (including mesh generation) obtained
at Notre Dame on a single node as a function of core count.
The data points show the mean total execution time for each core count, and the whiskers show $\pm$ two standard deviations over $10$ runs of each size, approximating a $0.95$ confidence interval. With $64$ cores, the average total time required to complete the simulation is $420.39$ seconds with a standard deviation of $36.29$ seconds. 
All three systems provided similar performance, validating the portability approach's effectiveness across HPC infrastructures.

\vspace{-0.05in}
\subsection{End to End Performance}
\label{sec:eval:e2e}

The end-to-end performance of \XGFABRIC\ is dominated by the time to prepare, queue, and execute the CFD simulation.  The telemetry data needed to generate the input files needed by OpenFOAM is available every $300$ seconds and requires approximately $200$ milliseconds to transfer from the 5G network at UNL to the head node of the cluster at ND via the data repository at UCSB (cf. Table~\ref{tab:cspot-message}).

If a $64$ core machine were dedicated to this application, it could sustain a rate of one simulation produced approximately every $7$ minutes.  However, these simulations are retrospective.  That is, they simulate the conditions that existed at the time just before the simulation was initiated -- not when it has completed.  Recall that the Laminar program that can trigger a statistical change requires $30$ minutes of telemetry data.  Thus, with $64$ cores, \XGFABRIC\ is able to generate a CFD simulation that is valid for a minimum of $23$ minutes (the $23$ minutes remaining after the $7$ minutes of simulation completes) up to the next change in wind speed.

However, making use of a shared computing facility
also results in queueing delay, which varies with
the offered load, the requested machines, the system capacity, and the scheduling discipline.  During the course of this project, the queueing delay at Notre Dame varied from zero to 24 hours at various points, and other facilities were no different.  The Pilot controller
(Section~\ref{sec:arch:hpc}) is designed to sidestep
this by submitting a pilot placeholder in advance,
and then "activating" the pilot as needed to achieve
real-time response.

Note that multi-node execution (using MPI) does not generate a speedup for the total application.  The OpenFOAM computation, itself, runs fastest on $2$ nodes, each with $64$ cores.  However, the total application (which includes both input-file generation and output postprocessing) slows down (despite the faster OpenFOAM time) when executed on more than one node.  However, \XGFABRIC\ is able to deploy the application to the best configuration possible given its performance needs, end-to-end.

%% file: conclusion.tex
\section{Conclusions and Future Work}
\label{sec:conclusion}

\XGFABRIC\ is a new, full-stack software platform designed to use 5G, (and emerging 6G) networking technologies to deliver HPC ``in-the-loop'' -- real-time or near real-time distributed applications that require high-performance computing components.
This prototype demonstrates the ability to leverage 5G/6G connectivity in remote, low infrastructure settings
such as commercial digital agriculture.  It also demonstrates the ability to use a single software stack on all devices in an IoT deployment, at all device scales, including HPC machines.  Moving forward, we plan to extend \XGFABRIC\ in several important ways:  First, we will incorporate the ability to use the dynamic control mechanisms available for 5G to implement IoT-tailored slicing techniques as a way of optimizing remote network usage.  Second, we will develop the Pilot infrastructure
to tune resource allocations in order to better avoid
batch queueing delays.  Finally, we will exploit the
simulation results to perform real-time interventions
in the CUPS facility.

%% file: appendix.tex
\twocolumn[
{\begin{center}
\Huge
Appendix: Artifact Description/Artifact Evaluation
\end{center}}
]

\appendixAD

\section{Overview of Contributions and Artifacts}

\subsection{Paper's Main Contributions}

\begin{description}
\item[$C_1$] Designing an end-to-end system for coupling sensor networks with HPC facilities through Private 5G networks to enable real-time-digital agriculture simulation.
\item[$C_2$] Demonstrating the ability to leverage 5G/6G connectivity in remote, low infrastructure settings such as commercial digital agriculture.
\item[$C_3$] Demonstrating the ability to use a single software stack on all devices in an IoT deployment, at all device scales, including HPC machines.
\end{description}

\subsection{Computational Artifacts}

\begin{description}
\item[$A_1$] \href{https://doi.org/10.5281/zenodo.16696837}{Full Artifact Repository (DOI: 10.5281/zenodo.16696837)}
\end{description}
\begin{description}
\item[$A_2$] \href{https://github.com/UNL-CPN-Lab/xGFabric-SC25-5G-Testbed}{Full Artifact Repository (github.com/UNL-CPN-Lab)}
\end{description}

\begin{center}
\begin{tabular}{r p{2cm} p{3cm}}
\toprule
Artifact ID & Contributions Supported & Related Paper Elements \\
\midrule
$A_1$ & $C_1, C_3$ & Figures 3, 7 \\
$A_2$ & $C_1, C_2$ & Figures 4, 5, 6 \\
\bottomrule
\end{tabular}
\end{center}

\section{Artifact Identification}

\newartifact

\artrel

The artifact provided, ($A_1$), includes the source code for the simulations to ensure and verify the reproducibility of the artifact. It also includes the plotting tools used to generate the original figures in the paper. The experiments were designed to be run on the head node and a compute node of an HPC cluster.

\artexp

\textbf{Figure 3:}

The replication of Figure 3 produces a PNG file depicting the final result from the CFD simulation of the CUPS farm. The figure shows the simulation of the airflow around the farm, with the wind velocity represented by color gradients. The simulation is based on a 3D model of the farm.

\vspace*{1em}
\noindent\textbf{Figure 7:}

The replication of Figure 7 produces a plot depicting the mean total runtime of the OpenFOAM simulation with varying number of cores on a single compute node. The plot shows the mean total runtime for each total number of threads, with error bars representing the 2 standard deviations across multiple runs.

\arttime

The artifact contains the source code and data to reproduce the speedup curve results presented in the paper. 

\begin{enumerate}
    \item[1)] \textit{Artifact Setup:} Once downloaded and configured, the artifact can be executed in less than 10 minutes.
    \item[2)] \textit{Artifact Execution:} The execution time of the artifact can varying greatly depending on queue times, number of jobs submitted concurrently, and number of cores. In total, the execution of one run for Figure 3 took around 15 minutes. For Figure 7, the total execution time was around 13 hours (780 minutes) with no queuing delay.
    \item[3)] \textit{Artifact Analysis:} For Figure 3, there is no analysis to be done. For Figure 7, the analysis of the results and generation of the figures can be done in less than 5 minutes once all the experiments have been executed.
\end{enumerate}

\artin

\artinpart{Hardware}

Computation is executed on both head nodes and compute nodes. The compute node should have UGE as its batch scheduler. A front-end node with display environment variables is required for rendering the CFD simulations.

\artinpart{Software}

The artifact should be executed on a Linux based machine with Bash, Python, and Pip installed. 
    
\artinpart{Datasets / Inputs}

The data for the artifact is provided in a zip file in the directory where it is used. The data includes the all necessary OpenFOAM files.

\artinpart{Installation and Deployment}

The first thing that the user needs to do is to access the head node of an HPC cluster with their display environment variables passed through via SSH. This is accomplished by adding the ``-Y'' flag when connecting to the front-end (e.g., ssh -Y user@HPC). Next, the user will install the necessary Pip packages listed in the requirements file.

\artcomp
The user will run the experiments by running either \textbf{``sh runme.sh -t=<number of threads>''} to run a single experiment or run them in batches with \textbf{``sh simulations.sh''}. If the user chooses the use the simulations file, they will first have to customize it with how many and which kind of runs they want. 

\artout
Once each experiments are complete, a user can run \textbf{``sh render.sh <name of experiment>''} for Figure 3 and/or \textbf{``python graphing.py''} for Figure 7. Notably, values are to be copied manually into the CSV file for replications of Figure 7. The values are to be retrieved from the result\_time logs. Each line of the CSV file with include: the experiment number and the total time that that the experiment took to run for the varying number of threads.

\newartifact

\artrel
The artifact ($A_2$) provided contains the source code used to build and deploy the 4G and the 5G network. It also includes a \texttt{data} folder containing all the experimental results, as well as plotting code used to reproduce Figures~4, 5, and 6 from the paper.

\vspace{1ex}
\noindent

\begin{samepage}
\noindent\textbf{Repository Structure:}

\begin{itemize}
    \item \texttt{build-4G-network/}: Contains source code used to build and deploy the private 4G network
    
    \item \texttt{build-5G-network/}: Contains source code used to build and deploy the private 5G network
    
    \item \texttt{data/}: Contains raw \texttt{iperf3} JSON output collected during all experiments, categorized by device, duplexing mode, and bandwidth setting.
    
    \item \texttt{visualize/visualize.ipynb}: Jupyter Notebook used to parse the experimental data and generate the plots shown in Figures~4, 5, and 6.
\end{itemize}

\noindent The repository supports full reproduction of the experiments and figures in the paper. Detailed setup and execution steps are provided in the \texttt{README.md}.
\end{samepage}

\artexp

\textbf{Figure 4:}

The replication of Figure 4 produces a PDF file depicting how uplink throughput performance varies across different bandwidths, duplexing modes, and user devices in a \textbf{single-user} scenario.

\vspace*{1em}
\noindent\textbf{Figure 5:}

The replication of Figure 5 produces a PDF file depicting how the uplink throughput performance varies across different channel bandwidths, duplexing modes, and user devices in a \textbf{two-user} scenario.

\vspace*{1em}
\noindent\textbf{Figure 6:}

The replication of Figure 5 produces a PDF file depicting how the results of slicing experiment conducted on the private 5G TDD network with a total bandwidth of 40MHz.

\arttime

The artifact contains the source code and data to reproduce the results presented in the paper. To generate the exact figures, the setup and execution can be skipped since the original data is included with the artifact.

\begin{enumerate}
    \item[1)] \textit{Artifact Setup:} Once downloaded, the artifact needs to be setup with docker. The docker process can take up to 30 minutes.
    \item[2)] \textit{Artifact Execution:} The execution time of the artifact can take up to 120 minutes to run all the experiments.
    \item[3)] \textit{Artifact Analysis:} The analysis of the results and generation of the figures can be done in less than 2 minutes once all the experiments have been executed.
\end{enumerate}

\artin

\artinpart{Hardware}

To run the experiments the user will need a Host Computer Running with 10th gen or later intel chip with at least 16GB RAM and an SDR (B210) connected over USB 3.0. 

\artinpart{Software}

The artifact should be executed on either a Ubuntu or MacOS based machine
with Bash, Python, Pip, and Docker installed

\artinpart{Datasets / Inputs}

The data for the artifact is provided in the data folder. The data can also be generated from the user.

\artinpart{Installation and Deployment}

Detailed setup and execution steps are provided in a README file for the 4G, 5G, and visualization folders.

\artcomp

The artifact can be executed by building the docker images, composing the docker images, and then running the \textbf{``data/script.sh''} file. At least 10-15 runs should be conducted for each device and network type for a sufficient sample size.

\artout

The outputs are JSON files that go into subfolder within the data folder. Each JSON file consists of metrics such as: host IP, destination IP, start time, end time, seconds, bytes, bits per seconds, etc.

\newpage
\appendixAE

\artin

The user needs to copy and paste the times from the \textbf{``result\_time''} log file into the corresponding CSV files located in \textbf{``data/data.csv''}. 

\artcomp

\textbf{For Figure 3:}

The user will copy the name of the folder of the completed run (e.g., cups\_structure\_25-07-29\_14\_07\_57) and use it in the following command: \textbf{``sh render.sh <name of folder>''}. This will generate a PNG file in the figures folder showing the completed CFD simulation of the farm. Note: the CFD output in Figure 3 was from a run with 64 threads.

\vspace*{1em}
\noindent\textbf{For Figure 7:}

The user will run at least 10 simulations for each number of threads for statistical significance. After all runs have finished and the times have been manually copied into the \textbf{``data/data.csv''} file, the user will run the \textbf{``graphing.py''} file. This will plot and save the graph as a PDF to the figures folder.

\artout
The user can run the plotting file located, \textbf{``graphing.py''}, to generate and save the figures.

\arteval{2}
\artin

When running the Jupyter Notebook, make sure that the NumPy and Matplotlib packages are installed. The data should already be available in the data folders.

\artcomp

Simply run all the cells of the Jupyter Notebook to generate the figures.

\artout

Figures 4, 5, and 6 are saved from the Jupyter Notebook as PDFs.

%% file: main.bbl

\begin{thebibliography}{23}


\ifx \showCODEN    \undefined \def \showCODEN     #1{\unskip}     \fi
\ifx \showDOI      \undefined \def \showDOI       #1{#1}\fi
\ifx \showISBNx    \undefined \def \showISBNx     #1{\unskip}     \fi
\ifx \showISBNxiii \undefined \def \showISBNxiii  #1{\unskip}     \fi
\ifx \showISSN     \undefined \def \showISSN      #1{\unskip}     \fi
\ifx \showLCCN     \undefined \def \showLCCN      #1{\unskip}     \fi
\ifx \shownote     \undefined \def \shownote      #1{#1}          \fi
\ifx \showarticletitle \undefined \def \showarticletitle #1{#1}   \fi
\ifx \showURL      \undefined \def \showURL       {\relax}        \fi
\providecommand\bibfield[2]{#2}
\providecommand\bibinfo[2]{#2}
\providecommand\natexlab[1]{#1}
\providecommand\showeprint[2][]{arXiv:#2}

\bibitem[Batur et~al\mbox{.}(2023)]%
        {Batur23Informs}
\bibfield{author}{\bibinfo{person}{Demet Batur}, \bibinfo{person}{Jennifer Ryan}, {and} \bibinfo{person}{Mehmet~C. Vuran}.} \bibinfo{year}{2023}\natexlab{}.
\newblock \showarticletitle{Dynamic Resource Sharing in Private {5G} Networks with Slicing}. In \bibinfo{booktitle}{\emph{INFORMS Annual Meeting}}. \bibinfo{address}{Phoenix, AZ}.
\newblock


\bibitem[Beckman(2025)]%
        {sage}
\bibfield{author}{\bibinfo{person}{Peter~H Beckman}.} \bibinfo{year}{2025}\natexlab{}.
\newblock \showarticletitle{Sage Grande: An Open Artificial Intelligence Testbed for Edge Computing and Intelligent Sensing}.
\newblock \bibinfo{journal}{\emph{NSF Award Number 2436842. Directorate for Computer and Information Science and Engineering}} \bibinfo{volume}{24}, \bibinfo{number}{2436842} (\bibinfo{year}{2025}), \bibinfo{pages}{36842}.
\newblock


\bibitem[Catlett et~al\mbox{.}(2022)]%
        {array-of-things}
\bibfield{author}{\bibinfo{person}{Charlie Catlett}, \bibinfo{person}{Pete Beckman}, \bibinfo{person}{Nicola Ferrier}, \bibinfo{person}{Michael~E Papka}, \bibinfo{person}{Rajesh Sankaran}, \bibinfo{person}{Jeff Solin}, \bibinfo{person}{Valerie Taylor}, \bibinfo{person}{Douglas Pancoast}, {and} \bibinfo{person}{Daniel Reed}.} \bibinfo{year}{2022}\natexlab{}.
\newblock \showarticletitle{Hands-on computer science: the array of things experimental urban instrument}.
\newblock \bibinfo{journal}{\emph{Computing in Science \& Engineering}} \bibinfo{volume}{24}, \bibinfo{number}{1} (\bibinfo{year}{2022}), \bibinfo{pages}{57--63}.
\newblock


\bibitem[Chen et~al\mbox{.}(2014)]%
        {openfoam}
\bibfield{author}{\bibinfo{person}{Goong Chen}, \bibinfo{person}{Qingang Xiong}, \bibinfo{person}{Philip~J Morris}, \bibinfo{person}{Eric~G Paterson}, \bibinfo{person}{Alexey Sergeev}, {and} \bibinfo{person}{Y Wang}.} \bibinfo{year}{2014}\natexlab{}.
\newblock \showarticletitle{OpenFOAM for computational fluid dynamics}.
\newblock \bibinfo{journal}{\emph{Notices of the AMS}} \bibinfo{volume}{61}, \bibinfo{number}{4} (\bibinfo{year}{2014}), \bibinfo{pages}{354--363}.
\newblock


\bibitem[Docker({[n.\,d.]})]%
        {docker}
\bibfield{author}{\bibinfo{person}{Docker}.} \bibinfo{year}{[n.\,d.]}\natexlab{}.
\newblock \bibinfo{howpublished}{\url{https://www.docker.com/}}.
\newblock


\bibitem[Ekaireb et~al\mbox{.}(2024)]%
        {laminar}
\bibfield{author}{\bibinfo{person}{Tyler Ekaireb}, \bibinfo{person}{Lukas Brand}, \bibinfo{person}{Nagarjun Avaraddy}, \bibinfo{person}{Markus Mock}, \bibinfo{person}{Chandra Krintz}, {and} \bibinfo{person}{Rich Wolski}.} \bibinfo{year}{2024}\natexlab{}.
\newblock \showarticletitle{Distributed dataflow across the edge-cloud continuum}. In \bibinfo{booktitle}{\emph{2024 IEEE 17th International Conference on Cloud Computing (CLOUD)}}. IEEE, \bibinfo{pages}{316--327}.
\newblock


\bibitem[Ettus({[n.\,d.]})]%
        {ettus}
\bibfield{author}{\bibinfo{person}{Ettus}.} \bibinfo{year}{[n.\,d.]}\natexlab{}.
\newblock \bibinfo{howpublished}{\url{https://www.ettus.com/}}.
\newblock


\bibitem[{farm-ng}(2024)]%
        {farm-ng}
{farm-ng} \bibinfo{year}{2024}\natexlab{}.
\newblock
\newblock
\newblock
\shownote{\url{https://farm-ng.com} [Online; accessed 5-Apr-2024]}.


\bibitem[Herath et~al\mbox{.}(1987)]%
        {df-strict}
\bibfield{author}{\bibinfo{person}{Jayantha Herath}, \bibinfo{person}{Toshitsugu Yuba}, {and} \bibinfo{person}{Nobuo Saito}.} \bibinfo{year}{1987}\natexlab{}.
\newblock \showarticletitle{Dataflow computing}. In \bibinfo{booktitle}{\emph{Parallel Algorithms and Architectures: International Workshop Suhl, GDR, May 25--30, 1987 Proceedings}}. Springer, \bibinfo{pages}{25--36}.
\newblock


\bibitem[Hintjens(2013)]%
        {zmq}
\bibfield{author}{\bibinfo{person}{Pieter Hintjens}.} \bibinfo{year}{2013}\natexlab{}.
\newblock \bibinfo{booktitle}{\emph{ZeroMQ: messaging for many applications}}.
\newblock \bibinfo{publisher}{" O'Reilly Media, Inc."}.
\newblock


\bibitem[Open5GS({[n.\,d.]})]%
        {open5gs}
\bibfield{author}{\bibinfo{person}{Open5GS}.} \bibinfo{year}{[n.\,d.]}\natexlab{}.
\newblock \bibinfo{howpublished}{\url{https://open5gs.org/}}.
\newblock


\bibitem[pysim({[n.\,d.]})]%
        {pysim}
\bibfield{author}{\bibinfo{person}{pysim}.} \bibinfo{year}{[n.\,d.]}\natexlab{}.
\newblock \bibinfo{howpublished}{\url{https://github.com/osmocom/pysim}}.
\newblock


\bibitem[RM530N-GL({[n.\,d.]})]%
        {5gDongle}
\bibfield{author}{\bibinfo{person}{RM530N-GL}.} \bibinfo{year}{[n.\,d.]}\natexlab{}.
\newblock \bibinfo{howpublished}{\url{https://www.waveshare.com/wiki/RM530N-GL}}.
\newblock


\bibitem[Rolshausen(2023)]%
        {crb-2023}
\bibfield{author}{\bibinfo{person}{Philippe Rolshausen}.} \bibinfo{year}{2023}\natexlab{}.
\newblock \showarticletitle{Prospects for Farming Citrus Under Protective Screening}.
\newblock \bibinfo{journal}{\emph{Citrograph}} \bibinfo{volume}{14}, \bibinfo{number}{4} (\bibinfo{year}{2023}).
\newblock


\bibitem[{Sage Grande Testbed}(2025)]%
        {sage-web}
{Sage Grande Testbed} \bibinfo{year}{2025}\natexlab{}.
\newblock \bibinfo{title}{{Sage Grande Testbed}}.
\newblock
\newblock
\newblock
\shownote{\url{https://sagecontinuum.org/docs/about/overview}[Online; accessed 3-Sep-2025]}.


\bibitem[SIM7600G-H-4GDONGLE({[n.\,d.]})]%
        {4gDongle}
\bibfield{author}{\bibinfo{person}{SIM7600G-H-4GDONGLE}.} \bibinfo{year}{[n.\,d.]}\natexlab{}.
\newblock \bibinfo{howpublished}{\url{https://www.waveshare.com/sim7600g-h-4g-hat.htm}}.
\newblock


\bibitem[Spertus and Dally(1991)]%
        {dataflow}
\bibfield{author}{\bibinfo{person}{Ellen Spertus} {and} \bibinfo{person}{William~J Dally}.} \bibinfo{year}{1991}\natexlab{}.
\newblock \bibinfo{booktitle}{\emph{Experiments with Dataflow on a General-Purpose Parallel Computer}}.
\newblock \bibinfo{publisher}{Massachusetts Institute of Technology, Artificial Intelligence Laboratory}.
\newblock


\bibitem[srsRAN Project({[n.\,d.]})]%
        {srsran}
\bibfield{author}{\bibinfo{person}{srsRAN Project}.} \bibinfo{year}{[n.\,d.]}\natexlab{}.
\newblock \bibinfo{howpublished}{\url{https://www.srsran.com}}.
\newblock


\bibitem[sysmoISIM SJA5({[n.\,d.]})]%
        {5gsim}
\bibfield{author}{\bibinfo{person}{sysmoISIM SJA5}.} \bibinfo{year}{[n.\,d.]}\natexlab{}.
\newblock \bibinfo{howpublished}{\url{https://sysmocom.de/}}.
\newblock


\bibitem[Turilli et~al\mbox{.}(2018)]%
        {pilot}
\bibfield{author}{\bibinfo{person}{Matteo Turilli}, \bibinfo{person}{Mark Santcroos}, {and} \bibinfo{person}{Shantenu Jha}.} \bibinfo{year}{2018}\natexlab{}.
\newblock \showarticletitle{A comprehensive perspective on pilot-job systems}.
\newblock \bibinfo{journal}{\emph{ACM Computing Surveys (CSUR)}} \bibinfo{volume}{51}, \bibinfo{number}{2} (\bibinfo{year}{2018}), \bibinfo{pages}{1--32}.
\newblock


\bibitem[{Waggled}(2025)]%
        {waggle}
{Waggled} \bibinfo{year}{2025}\natexlab{}.
\newblock \bibinfo{title}{{Waggle Software Stack}}.
\newblock
\newblock
\newblock
\shownote{\url{https://github.com/waggle-sensor}[Online; accessed 3-Sep-2025]}.


\bibitem[Wolski et~al\mbox{.}(2019)]%
        {wolski2019cspot}
\bibfield{author}{\bibinfo{person}{Rich Wolski}, \bibinfo{person}{Chandra Krintz}, {et~al\mbox{.}}} \bibinfo{year}{2019}\natexlab{}.
\newblock \showarticletitle{CSPOT: Portable, Multi-scale Functions-as-a-service for IoT}. In \bibinfo{booktitle}{\emph{ACM/IEEE Symposium on Edge Computing}}. \bibinfo{pages}{236--249}.
\newblock


\bibitem[Wolski et~al\mbox{.}(2025)]%
        {laminar-fmec}
\bibfield{author}{\bibinfo{person}{Rich Wolski}, \bibinfo{person}{Chandra Krintz}, {and} \bibinfo{person}{Markus Mock}.} \bibinfo{year}{2025}\natexlab{}.
\newblock \showarticletitle{Leveraging Dataflow as an Intermediate Representation for Portable Edge Deployments}. In \bibinfo{booktitle}{\emph{Proceedings of 10th International Conference on Fog and Mobile Edge Computing (FMEC 2025) -- to appear}}.
\newblock


\end{thebibliography}
